\documentclass[prb,twocolumn,showpacs,preprintnumbers,amsmath,amssymb,a4paper,superscriptaddress,floatfix,byrevtex]{revtex4}
\def\Vec#1{{\bf #1}}
\def\RE{{\rm Re}}
\usepackage{graphicx}
\usepackage{dcolumn}
\usepackage{bm}
\begin{document}
\preprint{$Revision: 5.0 $\hspace*{1cm}$Date: 2003-08-07 16:08:51+09 $}
\title{%
Immittance Matching for Multi-dimensional Open-system Photonic Crystals
}%
\author{Jun Ushida}
\email{ushida@cj.jp.nec.com}
\affiliation{%
Optoelectric Industry and Technology Development Association (OITDA)
}%
\affiliation{%
Fundamental Research Laboratories, NEC Corporation,
 34 Miyukigaoka, Tsukuba, 305-8501, JAPAN
}%
\author{Masatoshi Tokushima}
\affiliation{%
Fundamental Research Laboratories, NEC Corporation,
 34 Miyukigaoka, Tsukuba, 305-8501, JAPAN
}%
\author{Masayuki Shirane}
\affiliation{%
Fundamental Research Laboratories, NEC Corporation,
 34 Miyukigaoka, Tsukuba, 305-8501, JAPAN
}%
\author{Akiko Gomyo}
\affiliation{%
Fundamental Research Laboratories, NEC Corporation,
 34 Miyukigaoka, Tsukuba, 305-8501, JAPAN
}%
\author{Hirohito Yamada}
\affiliation{%
Optoelectric Industry and Technology Development Association (OITDA)
}%
\affiliation{%
Fundamental Research Laboratories, NEC Corporation,
 34 Miyukigaoka, Tsukuba, 305-8501, JAPAN
}%
\date{\today} 
%
\begin{abstract}
An electromagnetic (EM) Bloch wave propagating in a photonic crystal (PC) 
is characterized by the immittance (impedance and admittance) of the wave.
The immittance is used to investigate transmission and
 reflection at a surface or an interface of the PC.
In particular, the general properties of immittance are useful for 
clarifying the wave propagation characteristics.
We give a general proof that the immittance of EM Bloch waves on a plane
in infinite one- and two-dimensional (2D) PCs
is real when the plane is a reflection plane of the PC
and the Bloch wavevector is perpendicular to the plane.
We also show that the pure-real feature of immittance on a reflection
plane for an infinite three-dimensional PC is good approximation
based on the numerical calculations.
The analytical proof indicates that the method used 
for immittance matching is extremely simplified
since only the real part of the immittance function is needed for
analysis without numerical verification.
As an application of the proof, 
we describe a method based on immittance matching for qualitatively evaluating
the reflection at the surface of a semi-infinite 2D PC,
at the interface between a semi-infinite slab waveguide (WG)
and a semi-infinite 2D PC line-defect WG, and 
at the interface between a semi-infinite channel WG and
a semi-infinite 2D PC slab line-defect WG.
\end{abstract}
\pacs{42.70.Qs, 42.79.Gn}
\keywords{immittance of EM Bloch waves, impedance, admittance, photonic
crystal, photonic crystal slab, reflection plane,
 time reversal symmetry, reflection, transmission}
\maketitle
\section{Introduction}
Photonic crystals (PCs) are artificial materials whose dielectric
functions periodically vary.\cite{Yab,John,Ohtaka,Kosaka,Jo,Sakoda}
The primary characteristics of PCs are photonic band gaps and peculiar
dispersion relations, both of which enable the propagation of
electromagnetic (EM) waves in PCs to be controlled.
Recent interest in PCs has focused on 
``open-system PCs,''\cite{Jo,Sakoda,Noda,Mekis,Johnson,Stefanou,Yannopapas}
which are connected to other structures at their boundaries.
Typically investigated are a finite PC contacting another medium and
finite PC line-defect waveguides (WGs).
In open-system PCs, the energy of propagating EM waves is
transferred into or out of the PCs through the boundaries. This is in 
contrast to closed-system PCs, which are infinite in size and
therefore contain all the energy at any time.

An open-system PC can be divided into a set of subsystems --- the PC and the 
attached structures.  
Therefore, it should greatly facilitate designing open-system PCs
if each subsystem is independently designed and then all the subsystems
are assembled.
Such a method has become well established for electrical circuit design.
The behaviors of the electrical energy flows are predicted based on
electrical immittance (impedance and admittance).\cite{Heaviside_Woods}
Two electrical circuits can be assembled without energy reflection 
at their interface if the electrical immittances of both circuits match.
A complete electrical circuit system can thus be designed
after designing each subsystem independently.

EM wave media including open-system PCs can be regarded as EM wave
circuits, for which an analogous figure of merit, 
EM wave immittance, is similarly defined.\cite{Schelkunoff} 
It is a ratio of the electric field to the magnetic field of a propagating or
decaying wave in a single medium.
In practice, it has been possible to design one-dimensional (1D)
open-system PCs by using the EM wave immittance.
Particularly notable is that the renormalized Fresnel coefficient 
for a semi-infinite 1D PC with an arbitrary unit cell 
has been defined using the EM wave immittance 
for Bloch waves.\cite{Ushida_1d_arc} 
However, there is still difficulty in designing
multi-dimensional open-system PCs.
This is because the number of EM wave immittances
that should be considered at a point on an arbitrarily shaped boundary
increases with the number of structural dimensions.
Specifically, the immittance of an EM wave with a given polarization 
is expressed as one parameter for a 1D PC, two parameters for a 
two-dimensional (2D) PC, 
and six parameters for a three-dimensional (3D) PC.
Moreover, each parameter of an EM wave immittance is complex in general.
Therefore, excessive computational cost would be needed to match every 
immittance parameter between two multi-dimensional subsystems.

One way to simplify the matching problem is to reduce
the number of parameters by considering structural symmetry. 
Also, simplifying the parameters themselves is helpful, as has been 
suggested by Boscolo et al.\cite{Boscolo}  They pointed out that
a propagating wave in a PC line-defect WG has a pure-real immittance
on periodic symmetry planes normal to the direction of propagation.
Similar findings have been obtained
through analytical calculation of the impedance of
1D PCs\cite{Kawakami_2,Ushida_1d_arc} and 
through numerical calculation for 2D PCs.\cite{Noda_2}
In this article, we take another approach to simplifying
the matching problems --- qualitative optimization of individual subsystems.

This paper is organized as follows.
In Sec. \ref{sec:proof}, 
we first prove that pure-real immittance
is a common property of EM Bloch waves 
in 1D and 2D PCs, hence it can be used for any 1D or 2D open-system PC. 
We also prove that the pure-real property of the immittance 
is true for 3D periodic structures,  but under a limiting condition. 
Then, in Sec. \ref{sec:app}, we present 
three examples of designing  multi-dimensional open-system PCs.
This analysis can be made because the immittance defined by using
eigenmodes on the boundary of two semi-infinite regions
(a PC and its external structure) 
can be determined from the eigenmodes of 
an infinite PC and its external structure.
In Sec. \ref{subsec:2Dbulk}, we present  a method for qualitatively
evaluating the reflection at the surface of 
a semi-infinite 2D PC; 
the approximate availability of the perfect 
antireflection coating (ARC) method developed
for a semi-infinite 1D PC\cite{Ushida_1d_arc} to a semi-infinite 
2D PC is emphasized. These semi-infinite 2D PCs can be used
for analyzing various applications such as add/drop multiplexers, 
dispersion compensators, polarization filters, 
and image processors.\cite{Kosaka,Ohtera,Notomi} 
In Sec. \ref{subsec:2DPCWG}, 
we investigate a 2D line-defect WG that efficiently 
transmits EM energy to a channel WG. 
The semi-infinite 2D PC line-defect WG is 
a promising platform for ultra-compact photonic circuits.\cite{Johnson,Yamada}
In Sec. \ref{subsec:3d},  
we investigate the interface between a semi-infinite channel WG and
a semi-infinite 2D PC slab line-defect WG as a typical 3D problem.\cite{Miyai}
All the applications of the proof are presented using the concept of 
immittance matching.
Section \ref{sec:conclusion} devotes for conclusion.

\section{General proof for Immittance of Electromagnetic Bloch Waves}
\label{sec:proof}
We begin our description of the proof 
with a general expression of EM Bloch waves.
Assuming time harmonic EM waves, 
the general expression for an EM Bloch wave; \cite{Sakoda}
\begin{eqnarray}
\label{eqn:E}
 \Vec{E}(\Vec{r},t) &=& \Vec{u}_{\Vec{k}n} (\Vec{r})\ 
{\rm exp}(i\Vec{k}\cdot\Vec{r}-i\omega_{\Vec{k}n} t)\ , \\
\label{eqn:H}
 \Vec{H}(\Vec{r},t) &=& \Vec{v}_{\Vec{k}n}  (\Vec{r})\ 
{\rm exp}(i\Vec{k}\cdot\Vec{r}-i\omega_{\Vec{k}n} t)\ ,
\end{eqnarray}
where $n$ and $\Vec{k}$ stand for the band index and Bloch wavevector,
and functions $\Vec{u}_{\Vec{k}n}(\Vec{r})$ and 
	  $\Vec{v}_{\Vec{k}n}(\Vec{r})$ have lattice periodicity.
The immittance of a Bloch wave is closely related to the
energy flow conveyed by the wave; it is therefore useful to 
define a complex Poynting vector:
\begin{eqnarray}
\label{eqn:PoyntingVector}
\Vec{S}(\Vec{r}) = \frac{1}{2}\Vec{E}(\Vec{r},t)
\times\Vec{H}^*(\Vec{r},t) = \frac{1}{2}
\Vec{u}_{\Vec{k}n}(\Vec{r})\times
\Vec{v}^*_{\Vec{k}n}(\Vec{r})\ ,
\end{eqnarray}
where the real (imaginary) part of $\Vec{S}$ is 
the density of the time-averaged (reactive) power flow.\cite{Fano}

Here, we consider 
the time reversal state of Eqs. (\ref{eqn:E}) and (\ref{eqn:H}), which
is written as
\begin{eqnarray}
\label{eqn:t}
t'&=& -t\ {\rm and } \\
\label{eqn:Etr}
\Vec{E}^{\rm (tr)}(\Vec{r},t') &=& \Vec{E}(\Vec{r},-t)\ , \\
 \label{eqn:Htr}
\Vec{H}^{\rm (tr)}(\Vec{r},t') &=& -\Vec{H}(\Vec{r},-t)\ ,
\end{eqnarray}
due to the even (odd) parity of electric (magnetic) field
under time reversal operation.\cite{Jackson}
Eqs. (\ref{eqn:t})-(\ref{eqn:Htr}) can be reduced to\cite{Sakoda}
\begin{eqnarray}
 \label{eqn:omega}
 \omega_{-\Vec{k}n} &=& \omega_{\Vec{k}n}\, \\
 \label{eqn:u}
\Vec{u}_{-\Vec{k}n}(\Vec{r}) &=&\Vec{u}^*_{\Vec{k}n}(\Vec{r})\ , \\
 \label{eqn:v}
 \Vec{v}_{-\Vec{k}n}(\Vec{r}) &=&-\Vec{v}^*_{\Vec{k}n}(\Vec{r})\ .
\end{eqnarray}
Eqs. (\ref{eqn:omega})-(\ref{eqn:v}) express the time reversal
degeneracy of the spin 1 particle (photon).
Eqs. (\ref{eqn:omega})-(\ref{eqn:v}) will be used later.

%
  \begin{figure}
   \scalebox{1.15}{\includegraphics{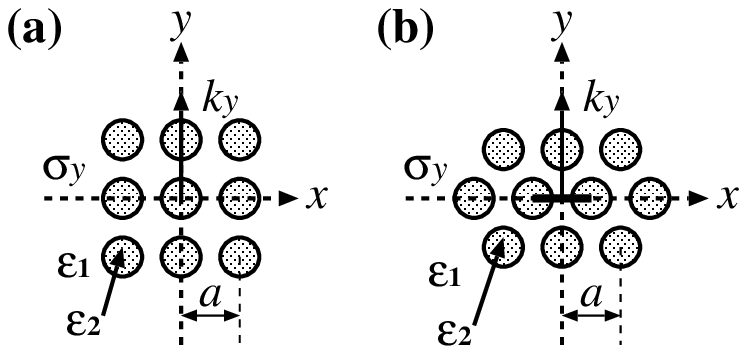}}
    \caption{Schematic illustration of typical 
infinite 2D PCs with reflection symmetry
   $\sigma_y$ (at $y=0$).
    (a) square lattice, (b) triangular lattice of circles with
   permittivity $\epsilon_2$ and lattice constant $a$ 
   in medium with $\epsilon_1$. 
    }
    \label{fig:Typical2DPC}
  \end{figure}
%

Next, we consider a Poynting vector described using
EM Bloch waves under a reflection operation in infinite 2D PCs.
Figure \ref{fig:Typical2DPC} illustrates typical infinite 2D PCs 
with reflection plane $\sigma_y$.
When the propagation direction of a Bloch wave is perpendicular to
this plane, the electric Bloch wave
$\Vec{E}(\Vec{r})$ is transformed by reflection operation into
\begin{eqnarray}
\label{eqn:Es1}
\hat{O}_{\sigma_y} \Vec{E}(\Vec{r})&=& 
IC_{2y}\left[
\begin{array}{l}
 u_{nk_y,x}\\
 u_{nk_y,y}\\
 u_{nk_y,z}
\end{array}
\right]
(x,-y,z)\ {\rm e}^{-ik_y y},\ \\
&=& 
\label{eqn:Es2}
\left[
\begin{array}{r}
 u_{nk_y,x}^*\\
-u_{nk_y,y}^*\\
 u_{nk_y,z}^*
\end{array}
\right]
(x,-y,z) \ {\rm e}^{ik_y y}, \ 
\end{eqnarray}
where $\hat{O}_{\sigma_y}$, $I$, and $C_{2y}$ stand for 
the $\sigma_y$ reflection operator, the inversion operator,  and 
the $C_{2}$ rotation about the $y$ axis. 
Note that the obvious factor ${\rm exp}(-i\omega_{\Vec{k}n}t)$ is
omitted in Eqs. (\ref{eqn:Es1}) and (\ref{eqn:Es2}).
Note also that this reflection operation ($\sigma_y$) 
is not the symmetry operation of the $\Vec{k}$-group at 
$k_y \hat{\Vec{e}}_y$, but
of the point group in an infinite 2D PC.
Similarly, magnetic Bloch wave $\Vec{H}(\Vec{r})$ is  transformed into
\begin{eqnarray}
\label{eqn:Hs1}
\hat{O}_{\sigma_y} \Vec{H}(\Vec{r})&=& 
\label{eqn:Hs2}
\left[
\begin{array}{r}
 v_{nk_y,x}^*\\
-v_{nk_y,y}^*\\
 v_{nk_y,z}^*
\end{array}
\right]
(x,-y,z) \ {\rm e}^{ik_y y}. \ 
\end{eqnarray}
In the derivation of Eqs. (\ref{eqn:Es2}) and (\ref{eqn:Hs2}),
we use Eqs. (\ref{eqn:u}) and (\ref{eqn:v}). 
The non-zero components of EM waves in the Cartesian coordinate system are 
$(H_x, H_y, E_z)$ for TM and $(E_x, E_y, H_z)$ for TE polarization.\cite{TMTE}

Accordingly, the Poynting vector described using the EM Bloch wave 
under reflection operation can be written as
\begin{eqnarray}
\label{eqn:Sy}
 S_y^{(\sigma_y)}(\Vec{r}) &=& 
\left\{
\begin{array}{rll}
    u^*_{nk_y,z} v_{nk_y,x}/2  & ({\rm for}\ {\rm TM})\\
   -u^*_{nk_y,x} v_{nk_y,z}/2  & ({\rm for}\ {\rm TE})\ .
\end{array}
\right.\ 
\end{eqnarray}
On reflection plane $y=0$,  
$S_y^{(\sigma_y)}(x,0,z)$ equals $S_y(x,0,z)$. Therefore, using 
Eqs. (\ref{eqn:PoyntingVector}) and (\ref{eqn:Sy}), we obtain
\begin{eqnarray}
\label{eqn:SSS}
&& S_y(x,0,z) = S_y^{(\sigma_y)}(x,0,z) \\
& \Leftrightarrow & 
\left\{
\begin{array}{rl}
\displaystyle
\frac{u_{nk_y,z}}{v_{nk_y,x}} = \left(\frac{u_{nk_y,z}}{v_{nk_y,x}}\right)^*
& ({\rm for}\ {\rm TM})\\
-
\displaystyle
\frac{u_{nk_y,x}}{v_{nk_y,z}} = -\left(\frac{u_{nk_y,x}}{v_{nk_y,z}}\right)^*
& ({\rm for}\ {\rm TE}) \\
\end{array}
\right. \\
\label{eqn:ImZ}
& \Leftrightarrow & 
\left\{
\begin{array}{rlr}
{\rm Im} (Z_{zx}) = 0
& ({\rm for}\ {\rm TM})\\
{\rm Im} (Z_{xz}) = 0
& ({\rm for}\ {\rm TE})\ ,\\
\end{array}
\right.
\end{eqnarray}
where $Z_{zx}\equiv u_{nk_y,z}/v_{nk_y,x}$, and 
$Z_{xz}\equiv-u_{nk_y,x}/v_{nk_y,z}$ at $(x,0,z)$.
\begin{flushright}
Q. E. D.
\end{flushright}
%
Eq. (\ref{eqn:ImZ}) means that the immittance 
is real in infinite 1D and 2D PCs on a plane  when the plane is a reflection
plane and the Bloch wavevector is perpendicular to the plane.
It also means that the reactive power flow is zero for the
Bloch wave on the plane under consideration.

For the case of an infinite 3D PC, Eq. (\ref{eqn:SSS}) can be
written as
\begin{eqnarray}
\nonumber
u_{nk_y,z}v_{nk_y,x}- u_{nk_y,x}v_{nk_y,z}&=&  \\ 
\left(u_{nk_y,z}v_{nk_y,x}\right.&-&\left. u_{nk_y,x}v_{nk_y,z}\right)^* \ .
\end{eqnarray}
Therefore, an additional reflection plane is required for removing
the imaginary part of the impedance. When there is reflection plane
$\sigma_z$ at $z=0$, some components of the EM waves vanish on the
plane. They are called TM- or TE-like based on the analogy to infinite 2D PCs.
\cite{Johnson_slab} In this case, the impedance is real
on the intersection line of the $y=0$ and $z=0$ planes.

%
  \begin{figure}
    \scalebox{.45}{\includegraphics{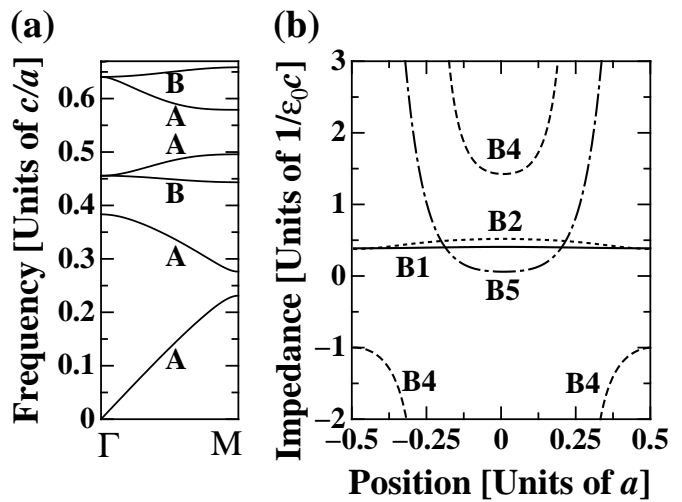}}
    \caption{(a) Dispersion relation of TM wave in $\Gamma$M
   direction for infinite 
   2D PC with triangular lattice of air holes ($ \epsilon_2=1$).
    Permittivity $\epsilon_1$ is 11.9025 and hole radius is
   0.4335$a$, where $a$ is lattice constant.
   Index ``A'' (``B'') indicates the coupled (uncoupled) 
   mode.\cite{Robertson1,Robertson2,Sakoda2}
    (b) Normalized impedance of Bloch waves 
    with $k_y=2\overline{\Gamma {\rm M}}/3$ at $y=0$.
    We use index B1-B5 (band number) to label the modes in order of
   increasing frequency. 
    Only Bloch waves that coupled with an external plane wave are shown.
    }
    \label{fig:ex1}
  \end{figure}

\section{Applications of pure-real property of immittance}
\label{sec:app}
As a typical application of the above proof, we present a 
method reducing a reflection loss at the surface of a semi-infinite
2D PC, a semi-infinite 2D PC line-defect WG, and 
a semi-infinite 2D PC slab line-defect WG
based on the concept of the immittance matching. 
The method for the immittance matching is extremely simplified
since only the real part of immittance function 
for an infinite PC is needed for the analysis.
The reflection coefficient of a semi-infinite PC 
should be determined via Bloch wave expansion 
in the semi-infinite PC and
the expansion should be described by all the eigen modes of an infinite
PC (i.e., including decaying waves).\cite{Minami} 
However, here we investigate only the impedance of propagation modes. 
This is another merit of our method, that is , 
it is not necessary to know all the eigenmodes in our method for 
immittance matching problem.
The information on the impedance of propagation modes 
can be used for qualitatively investigating
the reflection in various types of open system PCs.
\subsection{Semi-infinite 2D PC}
\label{subsec:2Dbulk}
The geometry of the first example is shown in Fig. \ref{fig:Typical2DPC} (b). 
The parameters of the PC is summarized in the caption of
Fig. \ref{fig:ex1}.

Figure \ref{fig:ex1} (a) illustrates 
the dispersion relation of the TM polarization wave 
for the infinite 2D PC in the $\Gamma$M direction.\cite{MPB}
Suppose that one considers the reflection at the surface of 
a semi-infinite PC made from the infinite PC cleaved at the plane $y=0$
shown in Fig. \ref{fig:ex1} (b). The incident wave is a normal incident 
plane wave propagating in vacuum.
Then, one should calculate the immittance of the Bloch wave at $y=0$ by using
the eigen mode of the infinite PC. The immittance of the Bloch wave is
generally complex and has lattice periodicity of the infinite 2D PC.
The impedance at $y=0$ calculated using propagating Bloch waves with 
$k_y=2\overline{\Gamma {\rm M}}/3$  is shown in Fig. \ref{fig:ex1} (b).
The illustrated region is indicated by the thick solid line 
in Fig.\ref{fig:Typical2DPC} (b).
The impedance is normalized by the impedance in vacuum.
Note that only the impedance of Bloch waves able to couple
with an incident plane wave (``coupled modes'') indicated by ``A''
in Fig. \ref{fig:ex1}(a) is plotted.
Since the spatial modulation of bands B1 and B2 is flatter than that of
bands B4 and B5 and the impedance is near the value in vacuum 
(i.e. impedance normalized by vacuum value is near one),
the reflection loss of bands B1 and B2 at the frequency 
is lower than that of bands B4 and B5 for
the normal incidence of a plane wave.
In addition, the flatness of the impedance enables us to reduce the
reflection loss by using an antireflection coating (ARC) method 
developed for semi-infinite 1D PCs.\cite{Ushida_1d_arc} 
In this case, since the plane $y=0$ is a
reflection plane, thickness $d$ and refractive index 
$n_2$ of the ARC can be estimated using 
\begin{eqnarray}
\label{eqn:D}
 D &\equiv& \frac{d}{\lambda_0/4n_2}= 1,3,5,\cdots\ \ \ {\rm and}\\
\label{eqn:n2}
n_2 &\simeq& \sqrt{n_1 \RE\left(Y_{zx}\right)/\epsilon_0 c}\ \ \  
{\rm at }\ \ x,y=0\ ,
\end{eqnarray}
where $D$ is the normalized thickness of the ARC,
$\lambda_0$ is a wavelength in vacuum, $n_1$ is the refractive index
of a semi-infinite medium,
and $Y_{zx} \equiv 1/Z_{zx}$ for TM polarization.
Accordingly, the reflection loss is easily removed by using 
Eqs. (\ref{eqn:D}) and (\ref{eqn:n2}). If a spatially flat impedance 
appears in the TE polarization case, Eq. (\ref{eqn:n2}) can be used by
replacing $Y_{zx}$ with $Y_{xz} \equiv 1/Z_{xz}$ 
to reduce the reflection loss of the TE polarization waves.

%
\subsection{Semi-infinite 2D Line-defect PC Waveguide}
\label{subsec:2DPCWG}

  \begin{figure}
   \scalebox{0.48}{\includegraphics{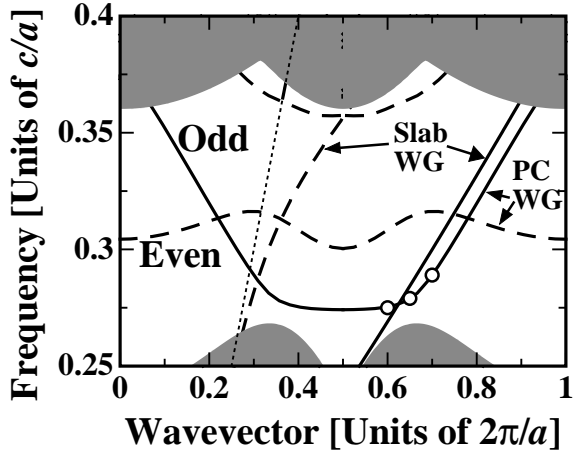}}
    \caption{Dispersion relation of infinite slab waveguide (WG) and 
   infinite PC WG. PC has triangular lattice of air holes.
 with radius $r= a/3$ in medium with $\epsilon_2=7.6176$.
   Even (odd) mode under reflection operation $\sigma_x$ bisecting WGs
   is shown as solid (broken) line.
   Circles show points equivalent to $K=$-0.30, -0.35, and -0.40,
   which are used to calculate impedances shown in
   Figs. \ref{fig:figure4}(a) and (b).
    }
    \label{fig:figure3}
  \end{figure}
%
The second example is related to the 
interconnection between a semi-infinite slab waveguide (WG) 
and a semi-infinite 2D PC line-defect WG, or ``PC WG.''
The connection between different WGs is important to realize
the ultra-minute photonic circuits.
Although some numerical investigation were made for similar 
structures,\cite{Yamada,Xu,Adibi,Adibi2,Mekis2,Mekis3,Miyai} we 
investigate it based on the concept of immittance matching.
The semi-infinite PC WG is made by removing a row of air holes 
in the $\Gamma$K direction 
from a semi-infinite 2D triangular lattice of air holes 
with radius $r=a/3$ and $\epsilon_2=1$ in a medium with
$\epsilon_1=7.6176$. The slab WG has width $\sqrt{3}a-2r$ and
permittivity $\epsilon=7.6176$ (the same as the $\epsilon_1$ of the PC).
The dispersion relation of the propagation modes of 
the infinite slab WG and infinite PC WG 
are shown in Fig. \ref{fig:figure3}.\cite{MPB} 
The polarization is TE; that is,
the magnetic wave is parallel to the slab and to the 
axis of air holes in the infinite 2D PC.\cite{TMTE}
Since the both of the slab WG and the PC WG have
reflection plane bisecting these WGs, the propagation mode has 
even (broken line) or odd (solid line) parity 
under reflection operation, as shown in Fig. \ref{fig:figure3}.
The shaded region indicates the projection band structure.
%
When K$=0.6\sim 0.9$, the dispersion relation of the infinite slab WG
is similar to that of the infinite PC WG. Such situation was called
$\Vec{k}$-matching (wavevector matching) by Mekis et al.\cite{Mekis2,Mekis3}
as a mechanism to reduce the reflection from the joint of
a distributed Bragg reflector WG and 
a 2D PC WG with square lattice of rods.
On the other hand, the exactly derived renormalized Fresnel
coefficients of a semi-infinite 1D PC did not include Bloch wavevector
explicitly.\cite{Ushida_1d_arc}
The analytical investigation of a relation 
between immittance matching and $\Vec{k}$-matching  will
be published elsewhere.\cite{Tokushima}
%
  \begin{figure}
   \scalebox{1.15}{\includegraphics{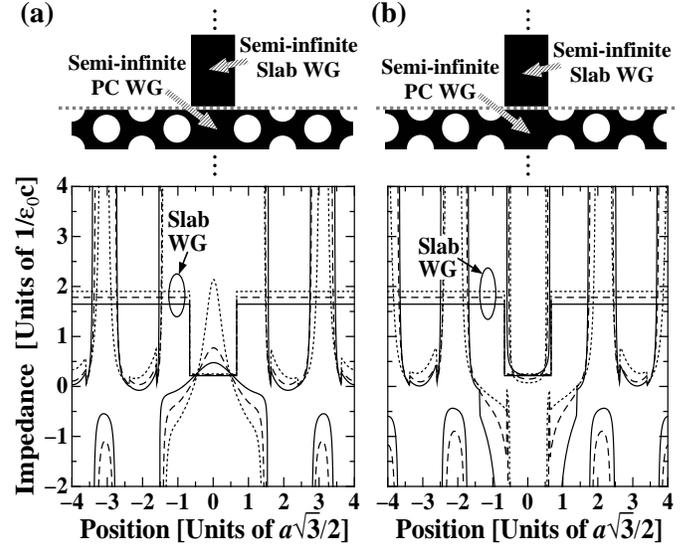}}
    \caption{Dependence of spatial distribution of impedance
    on wavevector and on surface of semi-infinite PC WG.
   Difference between Figs. \ref{fig:figure4} 
  (a) and (b) is the position of 
   the surface of semi-infinite PC line-defect WG 
   connected with semi-infinite slab WG, shown 
   as dotted line in diagram\cite{Miyai2} above each graph.
   Wavevector dependence is shown as solid (normalized
   wavevector $K=-0.30$), dashed ($K=-0.35$), and dotted
   ($K=-0.40$) lines. The rectangular impedance distribution is 
   impedances for slab WG with width $\sqrt{3}a-2r$.
    }
    \label{fig:figure4}
  \end{figure}
%

Here, we investigate the direct connection between a semi-infinite 
slab WG and a semi-infinite PC WG based on the immittance matching.
Figure \ref{fig:figure4} illustrates the dependence of the spatial
distribution of the impedance on the wavevector and on the surface of the
semi-infinite PC WG.
The difference between Figs. \ref{fig:figure4} (a) and (b)
is the position of the surface 
[shown as a dotted line in the diagram above each graph 
in Figs. \ref{fig:figure4}(a) and (b)] 
of the semi-infinite PC WG connected with the semi-infinite slab WG.
For the semi-infinite slab WG, 
the impedance is homogeneous inside and outside the slab
region, which shows a rectangular spatial distribution. In addition, 
the impedance at a surface is independent of the cleaved position of
the infinite slab WG.
For the semi-infinite PC WG,
the impedance depends strongly on the position of the surface of the
semi-infinite PC WG due to the spatial periodicity of the impedance in
the infinite PC WG.
Moreover, the impedance on a surface has a complex spatial distribution
on the surface.
Divergence appears where the magnetic wave vanishes. 
Divergence in a differential impedance 
appears at normalized position 
$X \equiv 2x/\sqrt{3}a =\pm2\pm2/3\sqrt{3}$,  
$\pm4\pm2/3\sqrt{3} \cdots$ 
for Fig. \ref{fig:figure4} (a), and at
$\pm1\pm2/3\sqrt{3}$, $\pm 3\pm2/3\sqrt{3}\cdots$ 
for Fig. \ref{fig:figure4} (b). The divergence is due to the discontinuity of
the electric wave perpendicular to the air holes at the surface of
the semi-infinite PC WG.
From the Poynting vector calculation, we can confirm that
the EM energy of these propagation modes concentrates
around the normalized position $X=0$.\cite{Boscolo} Therefore,
region $X=-1+2/3\sqrt{3}\sim1-2/3\sqrt{3}$ ($\sim$the width of the slab WG)
is quite important for analyzing the immittance matching
between the semi-infinite slab WG and semi-infinite PC WG modes. 
In this region, 
the impedance matching between the semi-infinite slab WG and 
the semi-infinite PC WG modes 
of Fig. \ref{fig:figure4} (b) is better than that of 
Fig. \ref{fig:figure4} (a).
Hence, the geometry of Fig. \ref{fig:figure4} (b) is better for 
reducing reflection loss. In addition, when the wavevector approaches
the 1st Brillouin zone boundary, the difference in the impedance
between the slab WG and the PC WG modes becomes noticeable. Therefore,
in a direct interconnection between the semi-infinite slab WG and 
the semi-infinite PC WG, 
the reflection loss of modes with $K=-0.4$ is higher than that with $K=-0.3$.

\subsection{Application to 3D structures
  --Semi-infinite 2D PC Slab Line-defect WG--}
\label{subsec:3d}
%
  \begin{figure}
    \scalebox{0.5}{\includegraphics{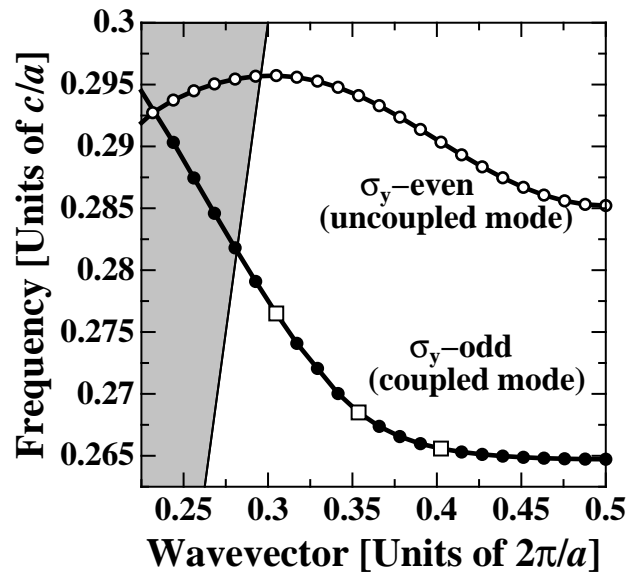}} 
    \caption{Dispersion relation of TE-like modes of
   2D PC slab line-defect WG. The slab consists of the triangular
   lattice of air holes (radius $r=0.29a$, thickness $h=0.6a$, refractive 
   index $n=3.4$, where $a$ is the lattice constant). The line defect is
   introduced by filling up a row of air holes. These parameters are
   the same as in Ref. \onlinecite{Chutinan}. 
   The 3D calculations are needed for this structure.\cite{MPB}
   The symmetry of the non-leaky 
   guided modes is indicated as $\sigma_y$-even or -odd, where
   $\sigma_y$ indicates the symmetry plane at $y=0$ in
   Fig. \ref{fig:figure6} (a) and (b). Squares stand for points
   executing impedance calculations shown in Figs. \ref{fig:figure6} (a) 
   and (b).
    }
    \label{fig:figure5}
  \end{figure}
\begin{figure*}
   \scalebox{0.7}{\includegraphics{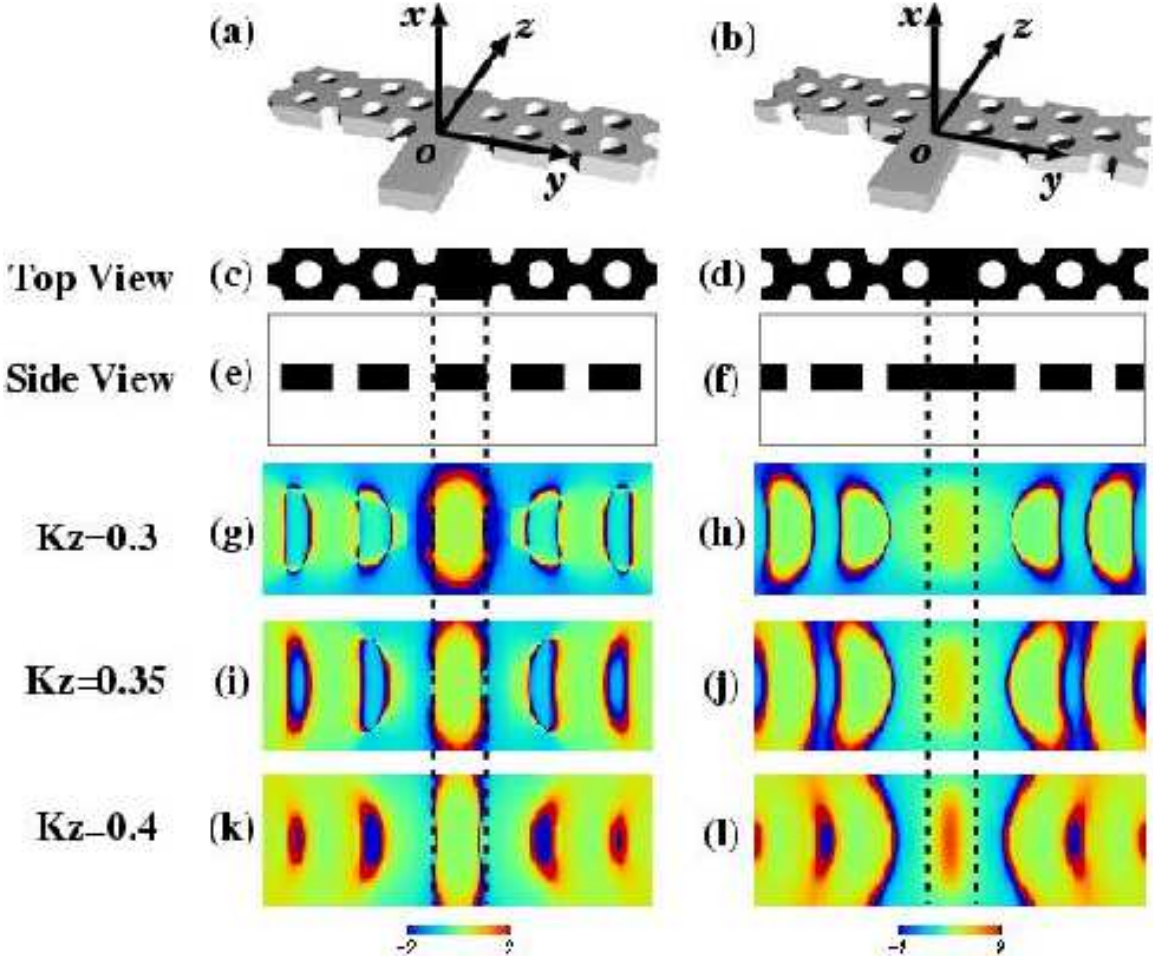}} 
    \caption{Dependence of spatial distribution of impedance
    on wavevector and on surface of semi-infinite 
   2D PC slab line-defect WG.
   Difference between Figs. \ref{fig:figure6} 
  (a) and (b) is the position of 
   the surface of semi-infinite PC slab line-defect WG 
   connected with semi-infinite channel WG ($z=+0$). 
   (c)-(f): top and side
   views of refractive index distribution. (g)-(l): 
   Dependence of impedance ${\rm Re}[Z_{yx}]$ (in units of $1/\epsilon_0 c$)
   on cutting surface and Bloch wave vector. 
   The dotted lines indicate the width of the channel WG.
    }
    \label{fig:figure6}
  \end{figure*}

%
In this section, we investigate the connection between a semi-infinite 
2D PC slab line-defect WG and a semi-infinite channel WG
based on the concept of immittance matching.
Schematic illustrations of two kinds of connections are shown 
in Figs. \ref{fig:figure6} (a) and (b). The structures are sandwiched
by two semi-infinite vacuum regions 
in the vertical direction ($x$-axis direction), 
hence they are 3D in nature. The 2D PC slab consists of
a triangular lattice of air holes. The line defect WG is introduced
into the slab by filling the air holes in a row. The detailed parameters
of these structures are given in the caption of Fig. \ref{fig:figure5}.
The width and thickness of the channel WG are $\sqrt{3}a-2r$ and $0.6a$, 
where $r$ is the radius of the air holes and $a$ is the lattice
constant.
%

%
%
Figure \ref{fig:figure5} illustrates the energy dispersion relation of
the TE-like modes of an infinitely long 2D PC line-defect WG.
Fundamental EM mode propagating in the channel WG can be coupled only with
the ``odd'' modes of 2D PC slab WG in the ideal case
[Mode symmetry is defined by the symmetry operation associated 
with the symmetry plane ($\sigma_y$ at y=0)
bisecting the 2D PC slab WG and the channel WG].
Therefore, we consider the odd mode only hereafter.

Even if the structures of the 2D PC slab WG and the channel WG were
fixed, there would still be ambiguity in determining
the structure of the connection --- the position of the cutting surface
in a unit cell of a 2D PC slab WG. This situation is the same as 
the 2D problem in Sec. \ref{subsec:2DPCWG}.
If we select a reflection plane as the cutting surface,
there are two possible choices, 
as illustrated in Figs. \ref{fig:figure6} (a) and (b) (at $z=+0$). 
To illustrate their difference more precisely, 
we show the top views ($x=0$ plane) and the side views ($z=0$ plane) of
the refractive index distribution in the unit cell used
for the supercell calculations\cite{MPB} in Figs. \ref{fig:figure6} (c)-(f); 
black indicates the regions with 
refractive index $n=3.4$, and white indicates the region with vacuum. 
The size of the supercell used in our 
calculation can be obtained from these figures.

When a 2D PC slab WG is terminated and connected with a
semi-infinite channel WG at $z=0$, 
the immittance distribution should be calculated at $z=+0$.
The analytical results presented in Sec. \ref{sec:proof} show that
the pure-real feature of immittance on a symmetry plane 
for a 3D structure is preserved along a particular line only. 
However, our numerical results indicate that
the imaginary part of immittance on a symmetry plane 
is much smaller than the real part.
In this situation, the imaginary part of the immittance on a symmetry
plane is negligible in qualitative immittance matching analysis.
Moreover, additional components of the immittance appear in the 3D problem.
This is because all components of an EM field 
cannot be treated separately.
The two possible combination of EM fields in the impedance calculation 
for Bloch waves propagating in the $z$ direction
are 
\begin{eqnarray}
\label{Zyx}
 Z_{yx}(x,y,+0) &=& -\frac{E_{k_z,y}(x,y,+0)}{H_{k_z,x}(x,y,+0)}\ , \\
\label{Zzy}
 Z_{xy}(x,y,+0) &=& \frac{E_{k_z,x}(x,y,+0)}{H_{k_z,y}(x,y,+0)}\ .
\end{eqnarray}
Since we use TE-like mode only in this analysis,
$Z_{xy}$ is much smaller than $Z_{yx}$, and this was confirmed by
our numerical calculations.
Therefore, in Figs. \ref{fig:figure6} (g)-(l) we illustrate
the real part of impedance distribution $Z_{yx}$ at $z=+0$. 
The regions of these plots are the same as those 
in Figs. \ref{fig:figure6} (e) and (f). 
The value of ${\rm Re}[Z_{yx}]$ is colored linearly.
Note that the region $|{\rm Re}[Z_{yx}]|\ge 2$ is colored red
or blue. Note also that the energy dispersion relation of the odd mode
shown in Fig. \ref{fig:figure5} has negative group velocity, so
the impedance is calculated by using time reversal states of EM fields.

The results presented in  Sec. \ref{subsec:2DPCWG}
clearly show that the structure in Fig. \ref{fig:figure6} (a) 
is more suitable for low-loss connections than that 
in Fig. \ref{fig:figure6} (b). This is supported by 3D calculations
of the impedance distribution around the WG region 
[$-(\sqrt{3}a-r)\le y \le \sqrt{3}a-r$ (between dotted lines 
in Fig.\ref{fig:figure6})].
The impedance distribution shown in Fig. \ref{fig:figure6} (g) 
is flatter than that in (h) at this region; 
meaning that the structure in (a) has lower
reflection than that in (b).

Next, we investigate their dependence on the Bloch wavevector, which 
is also shown in Figs. \ref{fig:figure6} (g)-(l).
When the length of the Bloch wavevector increases, the regions colored red
and blue expand. This means that 
the regions with $|{\rm Re}[Z_{yz}]|\ge 2$ expand, so 
the reflectance of EM waves increases.

\section{conclusion}
\label{sec:conclusion}
In conclusion, we presented a qualitative method for evaluating
the reflection of multi-dimensional semi-infinite PCs 
that is based on the concept of immittance matching.
Using the analytical investigation of
Poynting vectors defined using Bloch waves,
we presented a general proof that the imaginary part of
the complex immittance on periodic planes is zero 
in infinite 1D and 2D PCs when the plane is a reflection 
plane and the Bloch wavevector is perpendicular to the plane.
The infinite 3D PC case was also analyzed.
To show the usefulness of this proof,
we presented a qualitative method for evaluating 
reflection loss (a) at the surface of
semi-infinite 2D PCs, (b) at the interface between a semi-infinite 
slab WG and a semi-infinite PC line-defect WG, and (c) at the interface
between a channel WG and a semi-infinite 2D PC slab line-defect WG.
The first example showed approximate
applicability of the ARC method developed for semi-infinite 1D PCs 
to semi-infinite multi-dimensional PCs. 
The second example showed that the use of a connection with 
the geometry shown in Fig. \ref{fig:figure4} (b) reduces the reflection.
The third example showed that the pure-real feature of the immittance 
is a good approximation on a symmetry plane of 2D PC slab WG.

Note that reflection of PCs is sometimes discussed in relation with 
the group velocity or the density of states in the infinite PC.
While this can provide good physical intuition, we need to remember
that the reflectance strongly depends 
on the structure of the surface of a PC. The immittance of EM Bloch wave 
is essential for the discussion of the reflection phenomena for any 
``open-system'' PCs.
The general concept of immittance
matching was used for this theoretical prediction and is well suited
for reflection and transmission analysis
for a wide variety of ``open-system'' PCs.\\
%

\begin{acknowledgments}
\label{subsec:ack}
We thank Dr. Eiji Miyai and Professor Susumu Noda for showing us their work
(See, Ref. \onlinecite{Miyai2}) before publication. 
They also thank Mr. Arnold Fisher for his critical reading of manuscript.
One of authors (J.U.) thanks Dr. Steven G. Johnson 
for useful comments to the relation among adiabatic interconnections,
$\Vec{k}$-matching, and impedance matching at
private communication in Tsukuba, Japan on 19 Dec. 2002.
We thank Kazuo Nakamura and Jun'ichi Sone for their support.
This work was done as part of the ``Photonic Network Project''
under contract from the New Energy and Industrial Technology
Development Organization (NEDO) and
Focused Research and Development Project for the Realization of the
World's Most Advanced IT Nation, IT Program, MEXT.
%
%
\end{acknowledgments}

\end{document}